\documentclass[onecolumn,showkeys,showpacs,superscriptaddress,notitlepage]{revtex4-1}

\usepackage{graphicx}
\usepackage{amsmath,amsfonts,amssymb}
\usepackage{morefloats}
\usepackage{color}

\newcommand{\noi}{\noindent}
\newcommand{\be}{\begin{equation}}
\newcommand{\ee}{\end{equation}}

\begin{document}

\title[ ]{Social Pressure and Environmental Effects on Networks: A Path to Cooperation}

\author{Mar\'ia Pereda}\email{mpereda@math.uc3m.es}
\affiliation{Grupo Interdisciplinar de Sistemas Complejos (GISC), Departamento de Matem\'{a}ticas, Universidad Carlos III de Madrid, 28911 Legan\'{e}s, Madrid, Spain}

\author{Daniele Vilone}\email{daniele.vilone@gmail.com}
\affiliation{LABSS (Laboratory of Agent Based Social Simulation),
Institute of Cognitive Science and Technology,
National Research Council (CNR), 
Via Palestro 32, 00185 Rome, Italy}
\affiliation{Grupo Interdisciplinar de Sistemas Complejos (GISC), Departamento de Matem\'{a}ticas, Universidad Carlos III de Madrid, 28911 Legan\'{e}s, Madrid, Spain}

\begin{abstract}
In this paper, we study how the pro-social impact due to the vigilance by other individuals is conditioned by both environmental and evolutionary effects. To this aim, we consider a known model where agents play a Prisoner's Dilemma Game (PDG) among themselves and the pay-off matrix of an individual changes according to the number of neighbors that are ``vigilant'',  i.e., how many neighbors watch out for her behavior. In particular, the temptation to defect decreases linearly with the number of vigilant neighbors. This model proved to support cooperation in specific conditions, and here we check its robustness with different topologies, microscopical update rules and initial conditions. By means of many numerical simulations and few theoretical considerations, we find in which situations the vigilance by the others is more effective in favoring cooperative behaviors and when its influence is weaker.
\end{abstract}


\maketitle

\section{Introduction}

The emergence and survival of cooperative and, more in general, pro-social behaviors in nature and human societies has been one of the most debated issues in natural and social sciences for a long time~\cite{dar71,axe81,ham81}. Indeed, there is an apparent, yet paradoxical, contrast between the advantages of selfish strategies at the level of individuals, which should be expected to be mostly preferred by natural selection, and the ubiquitous presence of cooperation and altruism at the level of communities (not necessarily in humans)~\cite{may95}. In order to solve this problem, over recent decades many different mechanisms have been proposed~\cite{now92,now06,roc09,moy09}. What emerges from this great deal of study is that there is not a single universal mechanism that enhances cooperation against defection, but that different phenomena have a different explanation. In particular, if we limit our discussion to the pro-social behaviors in human communities, it has been demonstrated how indirect reciprocity~\cite{now98}, partner selection and punishment~\cite{ash96,and10} or gossip~\cite{gia16} can foster cooperative strategies in various situations.

Another factor which has shown its effectiveness in favoring human cooperation is the vigilance by others: more precisely, people tend to adopt more altruistic behaviors when they are observed by peers~\cite{tri71,ale87}, or even when they simply feel they are watched~\cite{bat06,ros07,oda15} ({\it Monitoring Hypothesis}). 
In~Ref.~\cite{per16}, a game-theoretical model able to describe this effect was presented. In particular, the effect of the vigilance was considered as a reduction of the temptation to defect in a Prisoner's Dilemma Game played in complex networks. As a result, the higher the level of vigilance, the higher the final degree of cooperation throughout a population. In that work, the behavior of the model was tested only in a few kinds of complex networks (essentially random and scale-free), with just one evolutionary rule (replicator) and always with the same initial conditions (completely random). \mbox{As we have }stressed above, the effect of the various mechanisms that determine the dynamics of these phenomena generally is not universal: for instance, the same topological structure can, in some cases, foster cooperation, or hinder it in different situations~\cite{roc09}. Therefore, in this paper, we aim to deepen and further clarify the results reported in Ref.~\cite{per16}, testing the robustness of its results by changing different aspects and parameters of the original model. In practice, we will focus on three factors: the topology (that is, the network on which the population evolves), the evolutionary algorithm (the rule following which the individuals adapt their strategies), and the initial conditions. \mbox{This is} important because, in the real world, communities live in different environments and evolve in different ways, \mbox{so that a test of t}his type allows for better evaluating the reliability of the model and the entirety of \mbox{its results}.

The paper is organized as follows: in the next section, we will define the model. Then, in Section~\ref{results}, we will present the results of the simulations, and, where possible, of some theoretical analysis. \mbox{Finally, in }Section~\ref{discuss}, we will discuss such results and sketch some perspectives.

\ 

\section{Model}

We consider a population of $N$ individuals interacting through an evolutionary Prisoner's Dilemma Game (PDG)
under vigilance pressure. The population is set on a given network, \mbox{which is }equivalent to assigning links
between the individuals that can interact directly: according to the distribution of links, the topology of the system will be different. Every player is characterized by a strategy, C
(cooperation) or D (defection), and, at each elementary time step, plays a round of the PDG with her neighbors,
and her neighbors do the same on their turn. After each interaction, an individual $i$ gets a payoff according to her payoff
matrix:
\be
\begin{tabular}{|c|c|c|}
\hline
\mbox{ } & {\bf C$_j$} & {\bf D$_j$} 
 \\ \hline
{\bf C$_i$} & $1$ & $0$
 \\ \hline
{\bf D$_i$} & $T_i$ & $P$ \\
\hline
\end{tabular}
\label{3b} 
\ee

\noi where $C_i,\ D_i$ are the strategies adopted by the player herself, and $C_j,\ D_j$ the strategies
used by the neighbor $j$; the {fitness} collected by $i$ in a single step of the dynamics will be the sum over all the payoffs collected with each neighbor. {Even though the averaged payoff per neighbor can also be used to define the fitness~\cite{ich14}, the total payoff allows for better singling out the role of the topology for the emergence of cooperation, and is more common in literature~\cite{axe81,roc09}.} Of course, to have a PDG, it must be $T_i>1\ \forall i$;
furthermore, {in order to reduce the parameters of the model, we fix the value of $P$ and} restrict to the weak Prisoner's Dilemma (wPDG), that is the case $P=0$~\cite{now92}. {Indeed, the wPDG has been often used in literature, since it makes the model simpler preserving its main features~\cite{roc09,moy09}.}

Moreover, every player can be either in a vigilant state, that is, controlling her neighbors' strategy, or not. Defining the variable $V_i$ which is equal to 0 if player $i$ is not vigilant, and equal to 1 if she is, \mbox{a non vigilant }individual can become vigilant following a Watts' threshold rule~\cite{per16,wat02}:
\be
\label{eq:vig}
{V}^{0 \rightarrow 1}_{i}({m}_{i},{k}_{i})= \left\{ \begin{array}{lcc}
             1 & if & {m}_{i}/{k}_{i} > {\theta}_{i}, \\
             \\ 0 & if &   {m}_{i}/{k}_{i} \leq {\theta}_{i} \ , \\
             \end{array}
   \right.
\ee

\noi where $m_i$ is the number of neighbors of the node $i$ that are already vigilant, $k_i$ is the degree of node $i$, and $\theta_i\in[0,1]$ the personal threshold of node $i$ above which she becomes vigilant. In this work, we consider this threshold constant and equal for every player: $\theta_i=\theta\ \forall i$. {In order to keep the model as simple as possible, we do not take into consideration any costs for becoming vigilant. This does not affect the generality of our results: as shown in reference~\cite{per16}, the cost of vigilance does not change qualitatively the behavior of the model, simply shifting possible transitions towards non-cooperative states for lower values of the temptation.}

The pressure due to the vigilance makes the temptation to defect effectively lower than in the absence of any external control: actually, it has already been demonstrated that people feel uncomfortable if they adopt anti-social behaviors just feeling observed~\cite{ros07,oda15}. In terms of the payoff matrix, \mbox{we can model} this phenomenon linking the temptation entry in the matrix~(\ref{3b}) to the number of vigilant neighbors:
\be 
T_i=b-\frac{m_i}{k_i}(b-1) \ , 
\label{vig_temp}
\ee

\noi where $b$ is the value of the temptation in the absence of vigilance. 

\subsection*{Evolutionary Rules} 
After all the individuals have played a round of the game, they update their strategies synchronously, according to a given rule. In this work, we have studied {two different update algorithms: unconditional imitation (UI), and a mixed update rule (MUR), inspired by Ref.~\cite{vil12}, and compared their performance to replicator (REP), the update rule used in Ref.~\cite{per16}}.

{\it Replicator} --- 
With REP, we proceed as follows. Let $s_i$ be the strategy the individual $i$ is playing, and~$\pi_i$ her payoff. With the proportional imitation rule, each individual $i$ randomly chooses one from her $k_i$ neighbors (individual $j$) and adopts her strategy with probability:

\begin{equation}
p^{t}_{ij}\equiv P \left\{s^{t}_{j} \rightarrow s^{t+1}_{i} \right\} = \left\{ \begin{array}{lcc}
(\pi^{t}_{j}-\pi^{t}_{i})/\Phi & if & \pi^{t}_{j} > \pi^{t}_{i}, \\
\\ 0 & if & \pi^{t}_{j} \leq \pi^{t}_{i}, \\
\end{array}
\right.
\end{equation}

\noi where $\Phi$ = max$(k_{i}, k_{j})$[max$(1,T)$$-$min$(0,S)$] so that $p^{t}_{ij} \in [0,1]$.

{\it Unconditional Imitation} --- with the UI rule, in order to evolve her strategy, every player imitates the one adopted by the neighbor that has obtained the best payoff, provided it is larger than her own (otherwise, nothing happens).

{\it Mixed update rule} --- in this case, with probability $q$, the player simply imitates the strategy of one of her neighbors picked up at random, and with probability $1-q$ evolves according the UI rule described above. While the REP rule is more representative of evolutionary phenomena in biology, this one describes better the dynamics underlying the decision making processes of human beings: therefore, it depicts more realistically social phenomena~\cite{vil12,vil14}.

In any case, whatever the update rule is, the strategies of the individuals are updated synchronously, {and, after the update, the payoffs of the players are set again to zero}. \mbox{Finally, after revising} their strategies {and payoffs}, players update their vigilance status, \mbox{according to the} rule given in Equation~(\ref{eq:vig}).

\ 

\section{Results}
\label{results}

We accomplished many simulations of the model defined in the previous section with different parameter values, topology, and update rules, in order to generalize the results presented in Ref.~\cite{per16}. In order to characterize and analyze the behavior of the model, we will consider the quantity $\langle\rho\rangle$, that is, the final average cooperator density{, measured after a transient of 100,000 generations and averaged over a time window of 100 generations, if the system has reached a stationary state defined by the slope of the average fraction of cooperators $\langle\rho\rangle$ being inferior to $10^{-2}$. If not, we let the system evolve subsequent time windows of 100 generations.} In this way, it will be easy to discern when the cooperation finally invades the system or is removed, or possible intermediate configurations.

All of the simulations presented here have been carried out with {systems of $N=1000$ individuals, large enough to consider negligible the finite size effects~\cite{san05}. Moreover, we confirmed the robustness of our results by accomplishing some simulations with smaller populations.}

{We will take into consideration basically monoplex networks, and} the topologies used in this paper are: (i) Erd\"{o}s-R\'{e}nyi (ER) random networks~\cite{erd60}; (ii) Barabasi-Albert (BA) scale-free networks~\cite{bar99}; (iii)~regular two-dimensional lattices (with absorbing boundary conditions); and (iv)~link-added small-world (LASW) random networks~\cite{wat98,vil11}. Unless explicitly indicated, the initial conditions are totally random, so that at the initial stage of the dynamics, on average, there are 50\% of cooperators; analogously, the initial vigilant players are also picked up at random: therefore, if only cooperators can be vigilant, we will have at the beginning the 25\% of vigilant cooperators. Otherwise, in Section~\ref{vig-def}, the initial vigilant individuals will be 50\% of the population, equally distributed among cooperators and defectors.

{Finally, we stress the fact that we aim to test the robustness of the outcomes presented in~\cite{per16}, \mbox{so that in }each of the next subsections, we will start usually from the original results and change only one feature of the model. Therefore, in Section~\ref{updr}, we will change, with respect to the analogous cases in~\cite{per16}, only the update rule, in Section~\ref{otop} the topology, and so on.}

\subsection{Influence of the Update Rule}
\label{updr}

{Here, we check how the behavior of the system changes by varying the way the individuals evolve their strategies, compared to Ref.~\cite{per16} (Section IIIA) where the REP rule is used, 
so we consider the same topology for comparison purposes, i.e., monoplex ER and BA networks with average degrees of $z=4$ and~$z=16$.}

\subsubsection{Unconditional Imitation}
\label{UIsec}

\begin{figure}[H]
\centering
\includegraphics[width=13cm,angle=0]{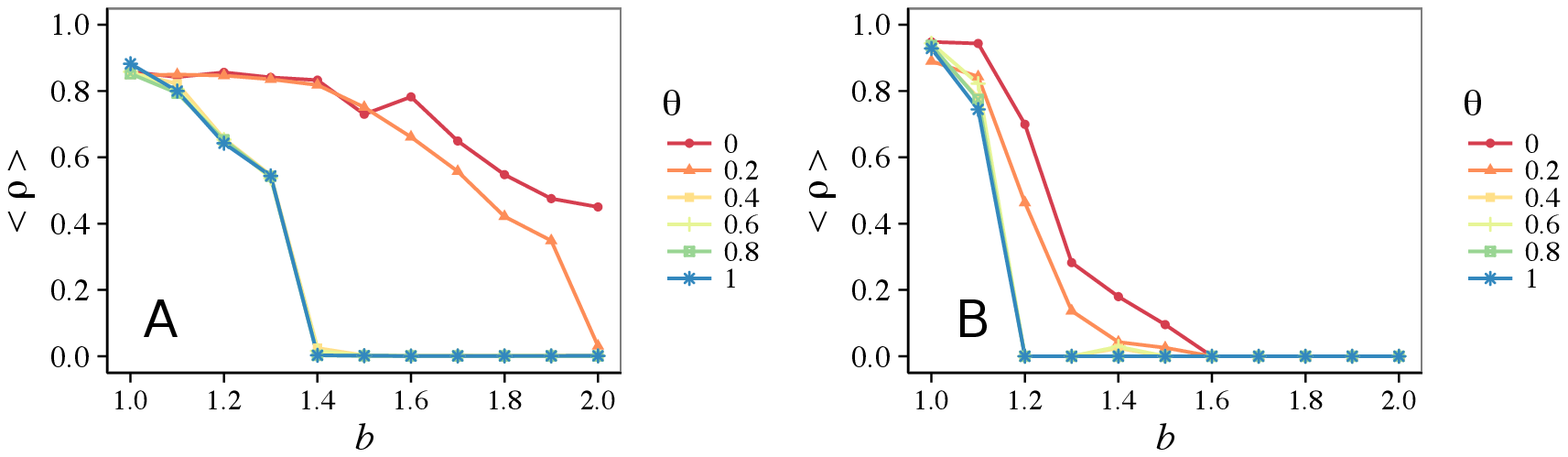}
\caption{Erd\"{o}s-R\'{e}nyi (ER) network.  \textbf{A}: $z=4$, \textbf{B} $z=16$; Unconditional Imitation (UI) evolution rule. Average final fractions of cooperators $\rho$ as a function of $b$ for different values of $\theta$.}
\label{uier}
\end{figure}

\begin{figure}[H]
\centering
\includegraphics[width=13cm,angle=0]{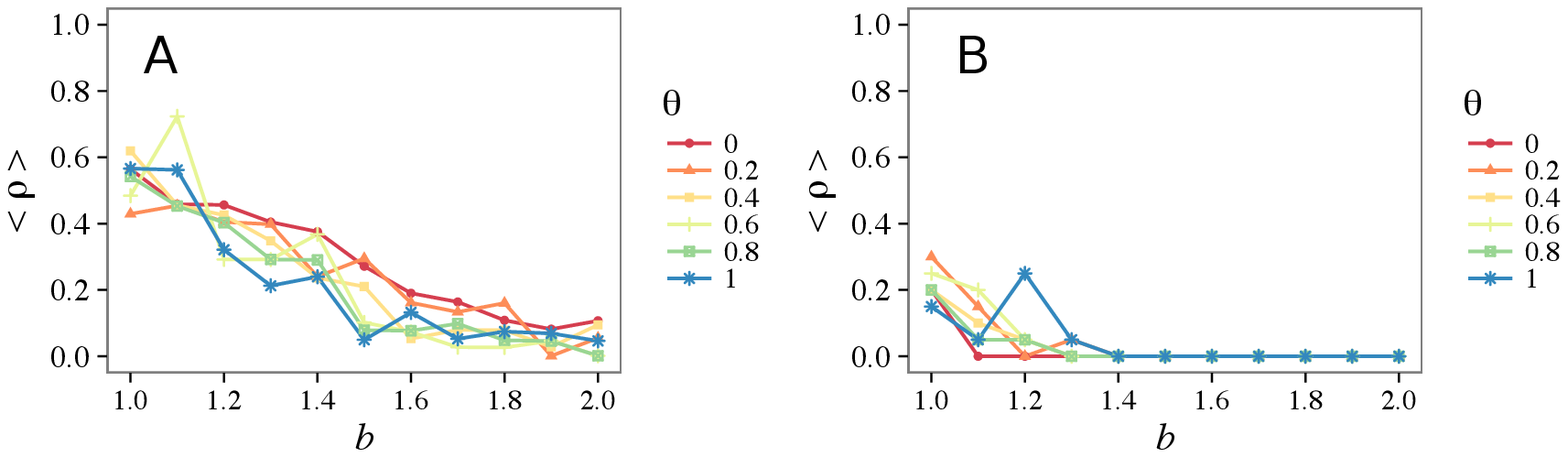}
\caption{Barabasi-Albert (BA) network.  \textbf{A}: $z=4$,  \textbf{B}: $z=16$; {UI evolution rule}. Average final fractions of cooperators $\rho$ as a function of $b$ for different values of $\theta$.}
\label{uiba}
\end{figure}

In Figure~\ref{uier}, the final average cooperation density as a function of the temptation $b$ is shown for different values of the threshold $\theta$ in the ER case, while, in Figure~\ref{uiba}, we report the same results for \mbox{a BA network}.

As it is easy to see, the cooperation is much more supported in ER topology than in BA. Comparing such results with the ones presented in Ref.~\cite{per16}, we notice that, with the REP rule, the cooperation \mbox{is favored} both in ER and BA networks. Therefore, we can conclude that the presence of hubs hinders the emergence of cooperative behaviors with a purely deterministic evolution algorithm,  i.e., a small amount of noise is necessary for cooperation to overcome this barrier. 

{The fact that, on scale-free topology, the unconditional imitation hinders cooperation} is further confirmed by taking into consideration a duplex BA-BA network, which is when the network of game dynamics and the one of vigilance dynamics are separated. {With replicator update, the system tends to fully cooperative final configurations up to large values of $b$~\cite{per16}; on the contrary, we verified numerically also that, with the UI rule, the final level of cooperation is very low already for $b\simeq1$.}

\subsubsection{Mixed Update Rule}

\begin{figure}[H]
\centering
\includegraphics[width=13cm,angle=0]{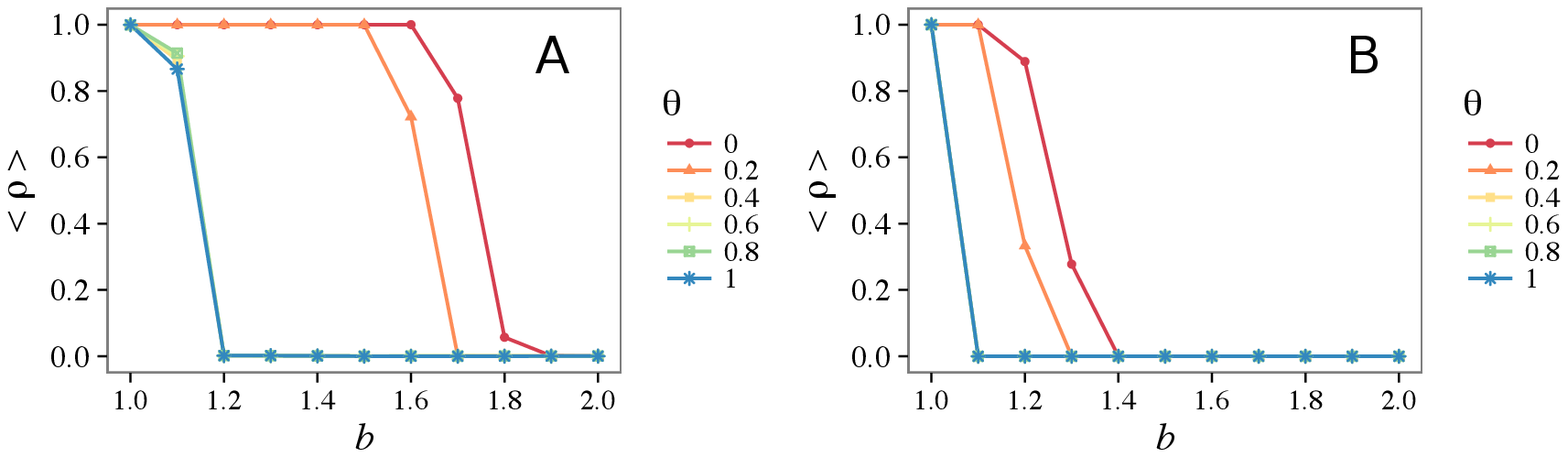}
\caption{Mixed update rule with $q=0.3$. ER monoplex network. \textbf{A}: $z=4$, \textbf{B}: $z=16$. \mbox{Average final} fractions of cooperators $\rho$ as a function of $b$ for different values of $\theta$.}
\label{ERmix_q03}
\end{figure}

\begin{figure}[H]
\centering
\includegraphics[width=13cm,angle=0]{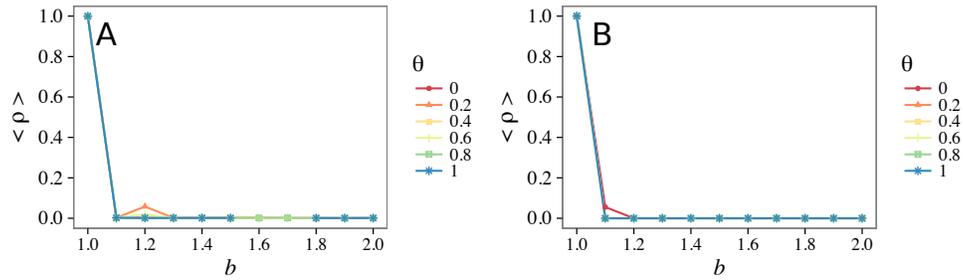}
\caption{Mixed update rule with $q=0.5$. ER monoplex network. \textbf{A}: $z=4$, \textbf{B}: $z=16$. \mbox{Average final} fractions of cooperators $\rho$ as a function of $b$ for different values of $\theta$. {Notice that all the graphics for the six different values of $\theta$ coincide almost everywhere.}}
\label{ERmix_q05}
\end{figure}

\begin{figure}[H]
\centering
\includegraphics[width=13cm,angle=0]{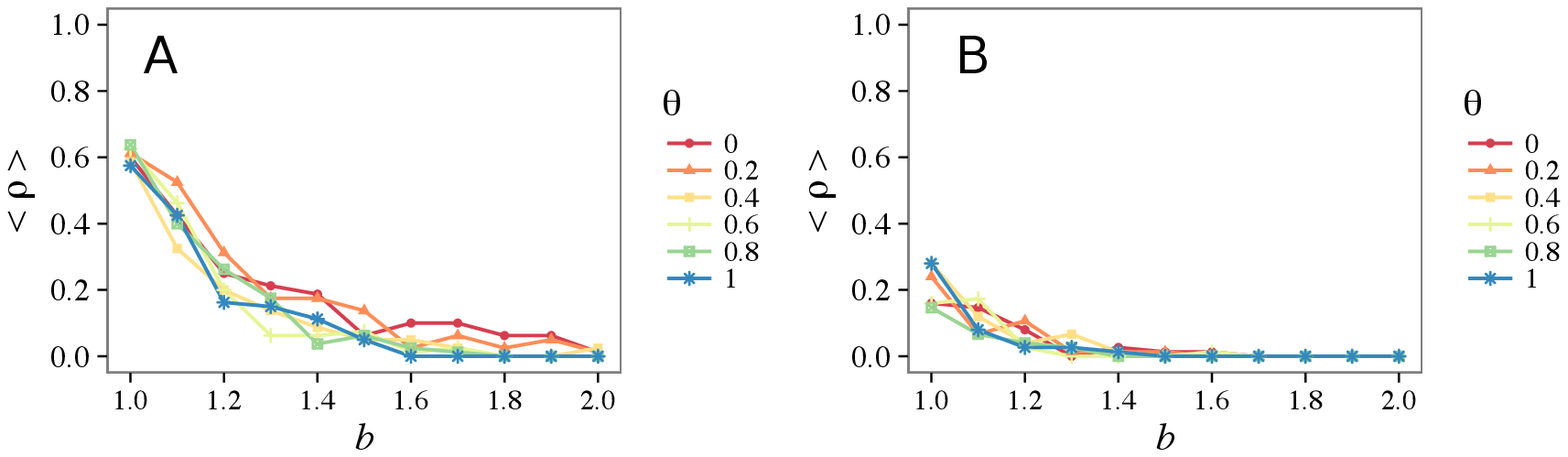}
\caption{Mixed update rule with $q=0.3$. BA monoplex network. \textbf{A}: $z=4$, \textbf{B}: $z=16$. \mbox{Average final} fractions of cooperators $\rho$ as a function of $b$ for different values of $\theta$.}
\label{BAmix_q03}
\end{figure}

\begin{figure}[H]
\centering
\includegraphics[width=13cm,angle=0]{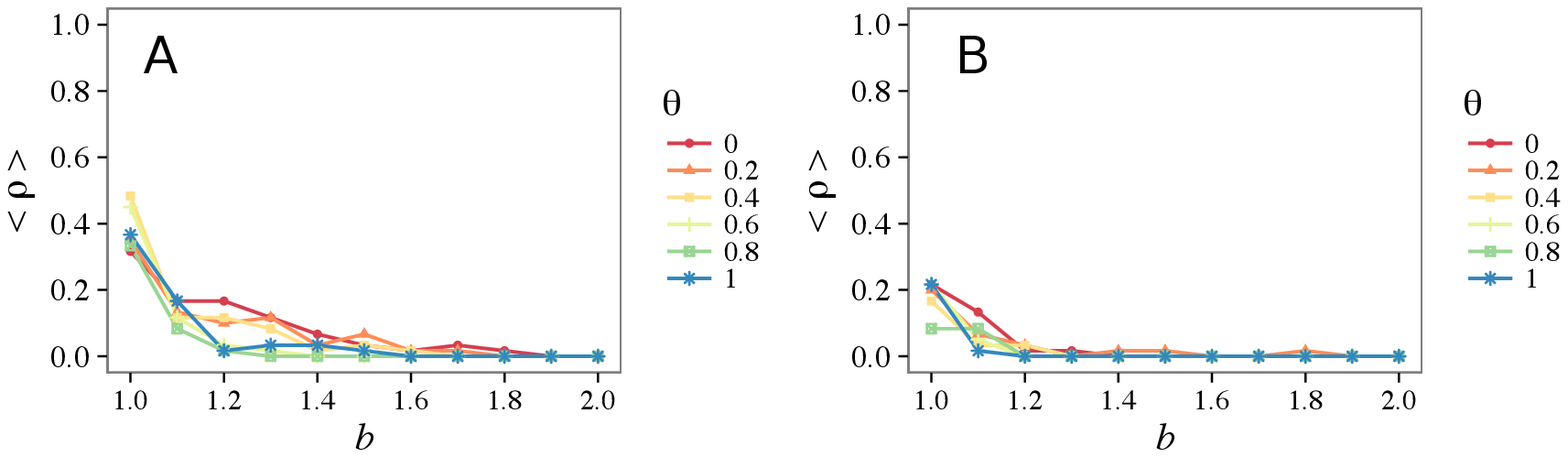}
\caption{Mixed update rule with $q=0.5$. BA monoplex network. \textbf{A}: $z=4$, \textbf{B}: $z=16$. \mbox{Average final} fractions of cooperators $\rho$ as a function of $b$ for different values of $\theta$.}
\label{BAmix_q05}
\end{figure}

We want now to check the robustness of the model with respect to the MUR rule, which is more realistic in the human interactions~\cite{vil14}. {Apart from the case of $q=0.3$ in ER networks (Figure~\ref{ERmix_q03}),} $\theta$~(vigilance) has no effect on cooperation, as shown in Figures~\ref{ERmix_q05},~\ref{BAmix_q03} and~\ref{BAmix_q05}. 
On the other hand, the value of the parameter $q$ ( i.e., the level of non-strategic imitation) does have effect: both for ER and BA. When the probability of following the non-strategic imitation rule is low ($q=0.3$) (Figures~\ref{ERmix_q03}~and~\ref{BAmix_q03}), we can find some levels of cooperation, but, with higher values (i.e., $q=0.5$) (Figures~\ref{ERmix_q05} and~\ref{BAmix_q05}), cooperation is hindered, as it happened in Section \ref{UIsec}.

It is worth stressing the fact that increasing the weight of the non-strategic imitation hinders the cooperation. This could be explained by considering that, by the UI rule, cooperators connected with other cooperators have a very high fitness and are surely imitated by a linked defector. To clarify this idea, let us consider a defector $j$ with four neighbors, among which there is only one cooperator $i$. Since cooperators tend to cluster, it is likely that the three defectors are connected to other defectors, getting in a single game round a fitness equal to four, whilst $i$ will be probably linked to three cooperators, gaining $3b$ (for the sake of simplicity, we assume that every individual has exactly four links). Thus, if $b>4/3$, player $j$ will definitely turn herself cooperator if evolved by the UI rule, while remain a defector with probability $3/4$ following the non-strategic update algorithm. Indeed, in { all Figures~\ref{ERmix_q03}--\ref{BAmix_q05}, we observe that the final cooperator density practically vanishes just around $b\approx1.2$--$1.3$}, coherently with the above considerations.

\subsection{Other Topologies}
\label{otop}

Up to now, we have considered the most classical examples of complex topologies, that is, \mbox{ER and BA} networks. Here, we aim to check the behavior of the model on topological structures with different features.
In particular, ER and BA networks differ mainly for the fact that, in the former, there are no hubs (nodes with much more connections with respect to the average), contrarily to what happens in scale-free BA networks~\cite{cal07}. In any case, both have a small diameter ( \mbox{i.e., the average} distance between two nodes picked up at random scales as the logarithm of the system size), \mbox{and a small clustering }coefficient (i.e., the probability that two neighbors of a third node are also neighbors is much smaller than 1). Therefore, it is worth considering networks with one or both diameters and clustering coefficients different from ER and BA networks. {As already hinted, we consider here only the replicator evolution rule, in order to compare the role of topology with the original results in~\cite{per16}.}

\begin{figure}[H]
\centering
\includegraphics[width=15cm,angle=0]{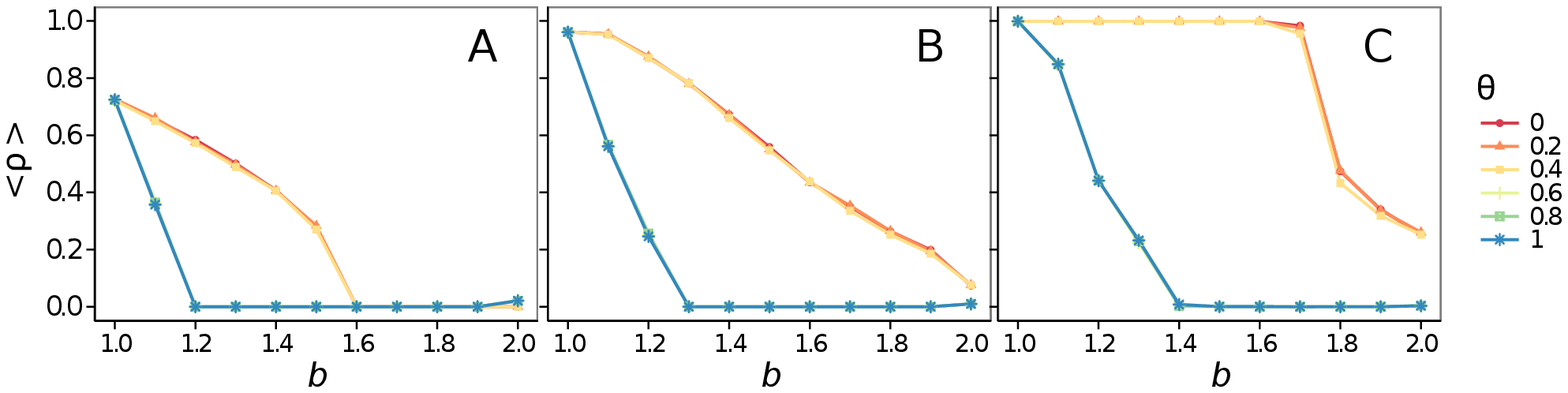}
\caption{Link-added small-world (LASW) network. {\bf A}: pure lattice; {\bf B}: lattice with $10\%$ of pure lattice number of links added; {\bf C}: lattice with $30\%$ of pure lattice number of links added. Average final fractions of cooperators $\rho$ as a function of $b$ for different values of $\theta$ --- Replicator (REP) evolution rule. In all of these graphs,\mbox{ the curves} for $\theta=0$ and $0.2$, and for $\theta=0.4,\ 0.6,\ 0.8$ and $1$, assume the same values.}
\label{uila}
\end{figure}

For this purpose, we took into consideration a Watts--Strogatz Small-World topology, \mbox{which has }the property of behaving locally as a regular lattice-like network (i.e., high clustering coefficient),\mbox{ but as a random }network globally (small diameter). Moreover, we built such a network following a different procedure from the one presented in Refs.~\cite{wat98,bar00}: starting from a regular square lattice of $N=1000$ nodes, each one with $z=4$ neighbors, we added links between non-connected nodes with a probability $p$, as in the LASW model defined in Ref.~\cite{vil11}. In this way, by tuning the parameter $p$, we can explore the lattice ($p=0$), and small-world ($0<p\lesssim 2z/N$) topologies.
Now,~as illustrated in Figure~\ref{uila}, we see how, in the lattice, the system cannot sustain cooperation (left graph), but increasing the density of short-cuts, the cooperation is mostly enhanced, even better than in ER topology (middle and right graphs). Interestingly, the results do not depend on $\theta$, apart from the fact that defection easily overcomes cooperation when $\theta=1$ already for small values of $b$.

\subsection{Different Initial Conditions}

A simple mean-field analysis of the model suggests that the outcome of the dynamics should also depend on the initial conditions, in particular on the initial distribution of the vigilant players. Actually, the vigilance can have an effective influence on the evolution of the system only if the vigilant individuals are enough to make the others vigilant too, following Equation~(\ref{eq:vig}). Now, considering \mbox{a mean-field} approach, the probability that an individual with $k$ connections has initially $m$ vigilant neighbors is

\be
\label{mf-1}
P(m;k) = \binom{k}{m}a_0^m(1-a_0)^{k-m} \ , 
\ee

\noi where $a_0$ is the initial density of the vigilant individuals. Then, the average density of vigilant neighbors at the beginning of the dynamics can be easily computed:

\be 
\label{mf-2}
\left\langle\frac{m}{k}\right\rangle = \sum_{m=0}^k\frac{m}{k}P(m;k)=a_0 \ . 
\ee

Therefore, the effect of vigilance should become noticeable for $\theta<a_0$: since we usually set initially half of cooperators as also being vigilant, we expect a transition from high cooperation to defection for $\theta$ larger than a critical $\theta^*$ such that

\be 
\label{mf-tstar}
\theta^*\approx\frac{\rho_0}{2} \, 
\ee

\noi where $\rho_0=0.5$ is the initial cooperator density. Of course, we also expect that the network structure changes at least partially this picture. In fact, the influence of the initial conditions is almost completely removed in non-trivial topologies, as we are going to show in the following.

\begin{figure}[H]
\centering
\includegraphics[width=6.5cm,angle=0]{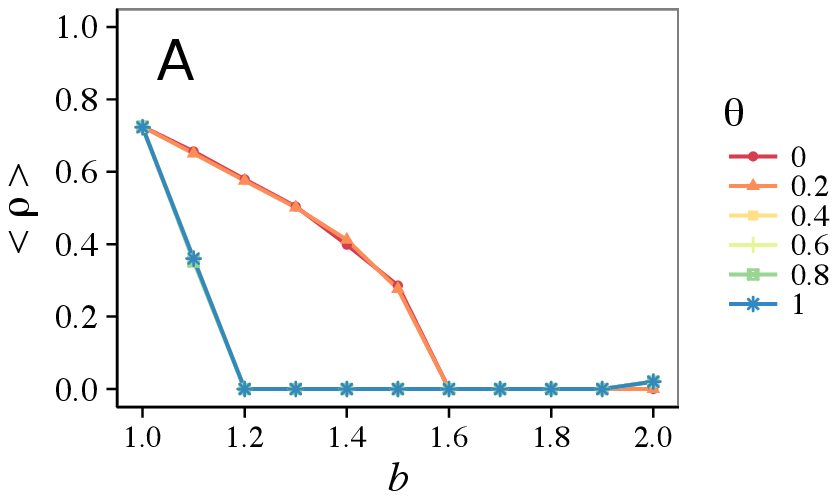} 
\includegraphics[width=6.5cm,angle=0]{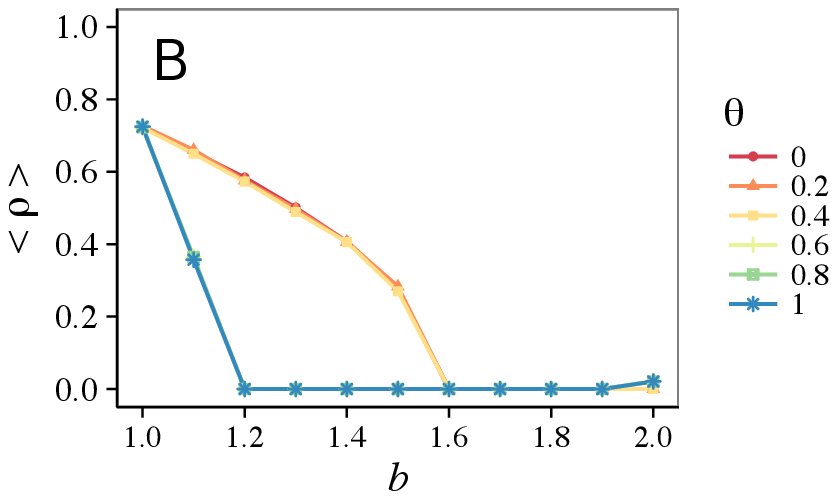}
\includegraphics[width=6.5cm,angle=0]{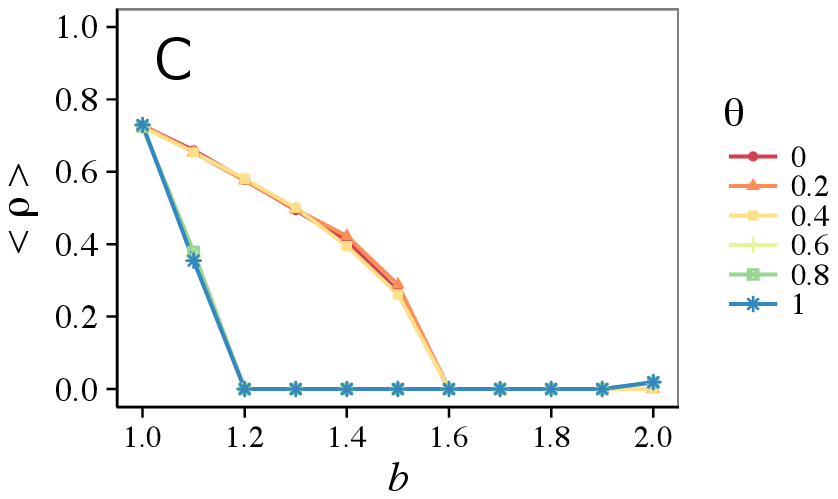} 
\includegraphics[width=6.5cm,angle=0]{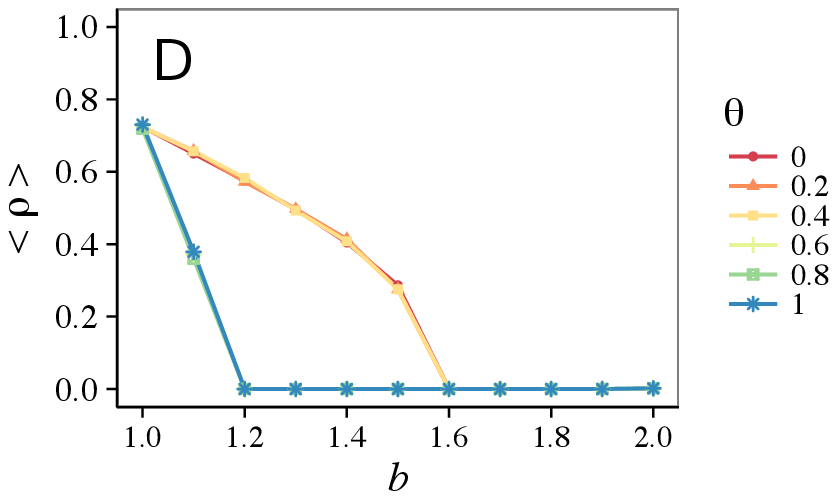} \\
\caption{Square lattice, different initial vigilant densities. {\bf A}: only one initial vigilant put in the middle of the lattice; {\bf B}: initial probability for each cooperator to be vigilant \mbox{equal to 0.001}; {\bf C}: initial probability to be vigilant equal to 0.05; {\bf D}: initial probability to be vigilant equal to 0.45. Average final fractions of cooperators $\rho$ as a function of $b$ for different values of $\theta$ --- {\mbox{REP evolution rule}. In the top-left graph, the curves from $\theta=0$ and $0.2$, and from $\theta=0.4$ assume the same values; in the remaining ones, the curves from $\theta=0,\ 0.2$ and $0.4$, and from $\theta=0.6$ assume the same values.}}
\label{initLA}
\end{figure}

In Figure~\ref{initLA}, we present the final cooperator distribution for a system on a square lattice evolving by the REP rule. As it is easy to realize, if the number and distribution of initial vigilant individuals is such that no other player can be activated, then there will be no effect of the vigilance and the cooperation vanishes already for small values of the temptation $b$. On the contrary, as the initial distribution allows, even through statistical fluctuations, that some inactive player can have enough vigilant neighbors to get activated, then the number of vigilant individuals soon increases and the system ends up in \mbox{a configuration} with a higher level of cooperation, independently from the initial number of \mbox{vigilant agents}.

\begin{figure}[H]
\centering
\includegraphics[width=13cm,angle=0]{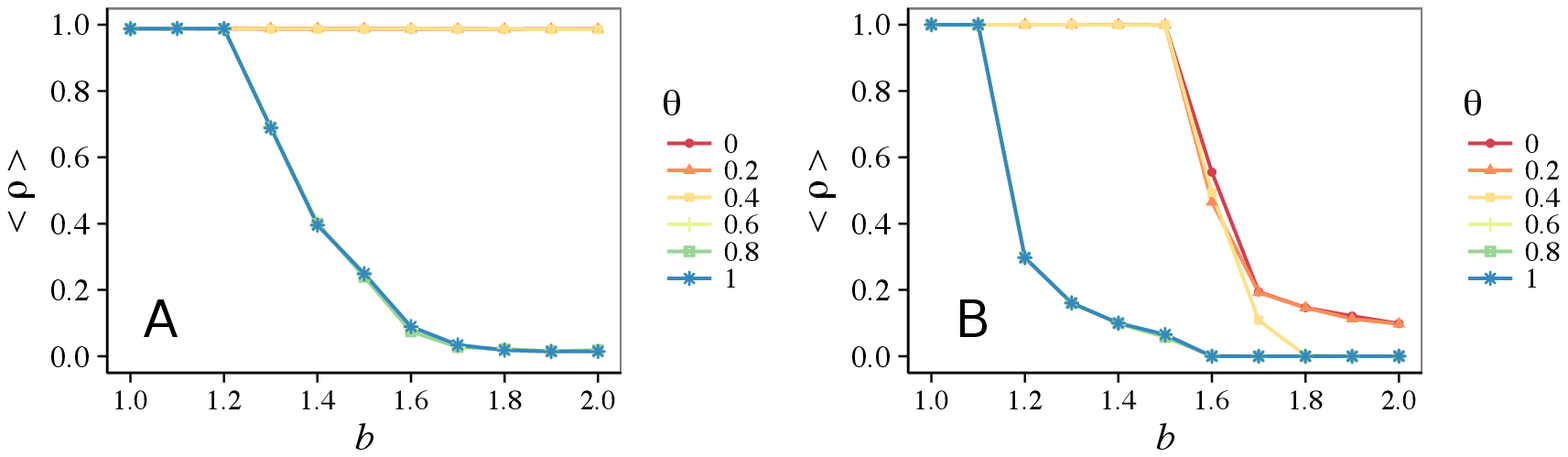} 
\includegraphics[width=13cm,angle=0]{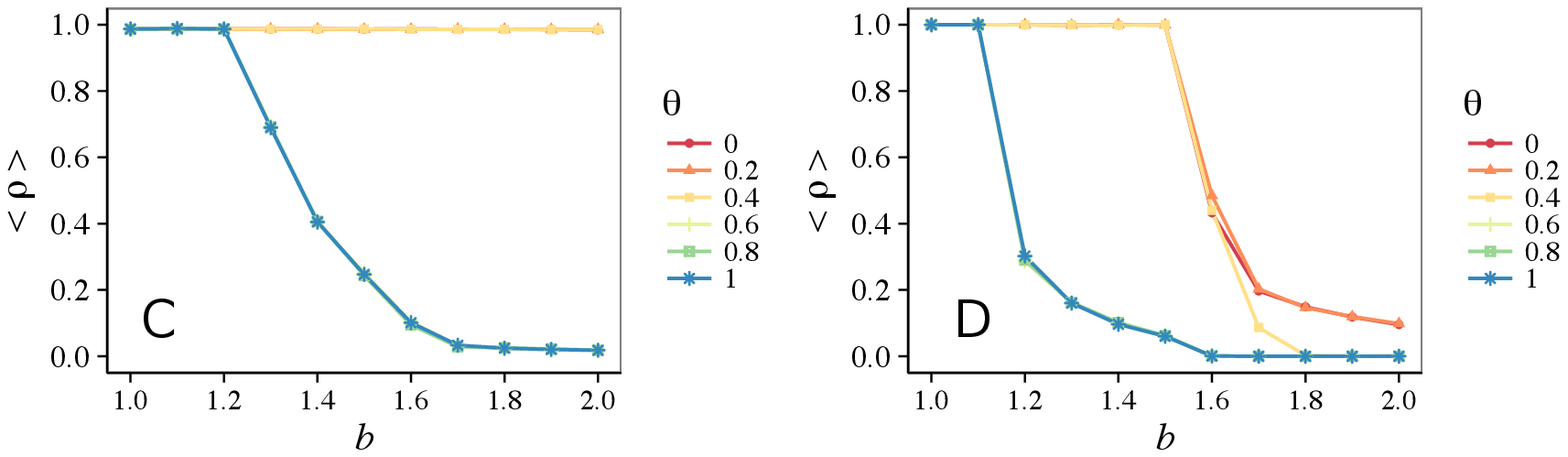} \\
\caption{ER network, different initial vigilant densities.\textbf{ A, C} graphs: $z=4$; \textbf{B, D} graphs: $z=16$; \textbf{A, B} figures: initial probability to be vigilant 0.05; \textbf{C, D} figures: initial probability to be vigilant 0.45. Average final fractions of cooperators $\rho$ as a function of $b$ for different values of $\theta${---REP evolution rule. In the left graphs, the curves from $\theta=0,\ 0.2$ and $0.4$, and from $\theta=0.6$ assume the same values; \mbox{in the remaining ones}, the curves from $\theta=0.6$ assume the same values.}}
\label{initER}
\end{figure}

\begin{figure}[H]
\centering
\includegraphics[width=13cm,angle=0]{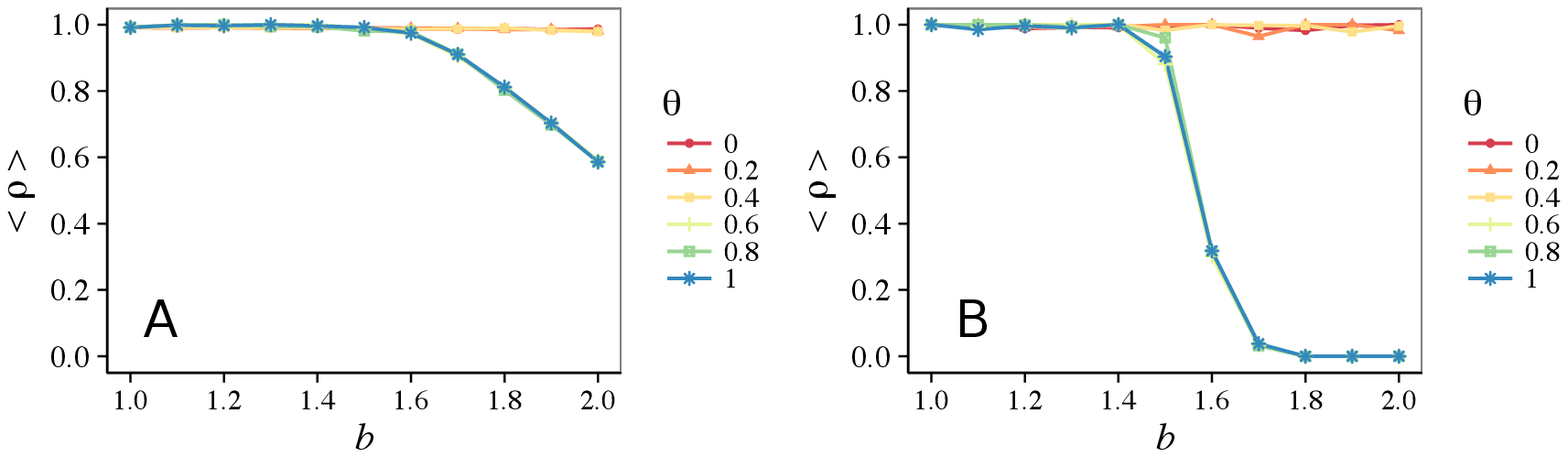} 
\includegraphics[width=13cm,angle=0]{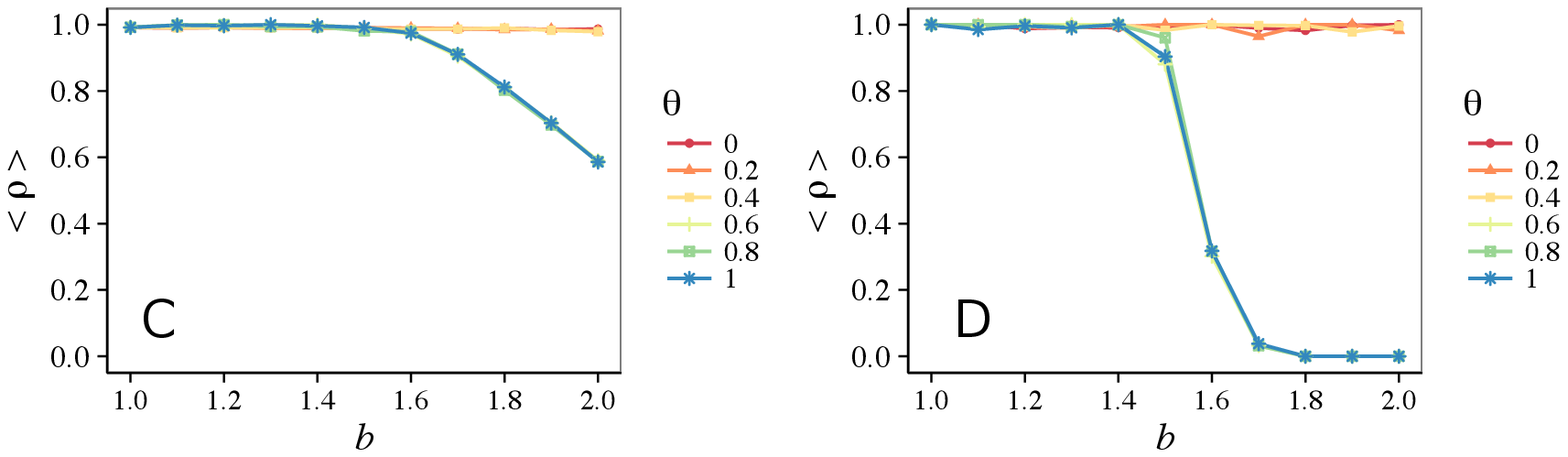} \\
\caption{BA network, different initial vigilant densities. \textbf{A, C} graphs: $z=4$; \textbf{B, D} graphs: $z=16$; \textbf{A, B} figures: initial probability to be vigilant 0.05; \textbf{C, D} figures: initial probability to be vigilant 0.45. Average final fractions of cooperators $\rho$ as a function of $b$ for different values of $\theta${---REP evolution rule. In the left graphs, the curves from $\theta=0,\ 0.2$ and $0.4$, and from $\theta=0.6$ assume the same values; \mbox{in the remaining ones}, the curves from $\theta=0.6$ assume the same values.}}
\label{initBA}
\end{figure}

This is true also in ER random networks, as shown in Figure~\ref{initER}: in the end, there is practically no effect of the initial vigilant density on the final fate of the dynamics. Indeed, as can be proven by comparing these results with the Figure~1a of the Ref.~\cite{per16}, $\langle\rho\rangle$ is always very close to the value of the case $a_0=0.25$, apart some slight differences.
This same picture holds for BA networks as well: also with this topology, the final level of cooperation does not depend on the initial distribution of the activated players, as reported in Figure~\ref{initBA}.

Therefore, we can finally state that the dynamics turn out to be robust with respect to varying the initial conditions, so that what has been presented in the previous subsections can be considered as general results with respect to the initial configuration of the system.

\subsection{Case of Vigilant Defectors}
\label{vig-def}

Until now, we have set that only cooperators can also be vigilant players. In fact, in a PDG, defectors also have interest in being connected with cooperators, so it is plausible to also consider a situation where someone who is not a cooperator can be vigilant. In practice, in human interactions, those who also adopt anti-social behaviors can force the others to behave fairly~\cite{gia16,dun04,gia12}. 

Therefore, we considered the case in which every player, independently from the fact that she is either a cooperator or a defector, can be a vigilant one. 

\begin{figure}[H]
\centering
\includegraphics[width=13cm,angle=0]{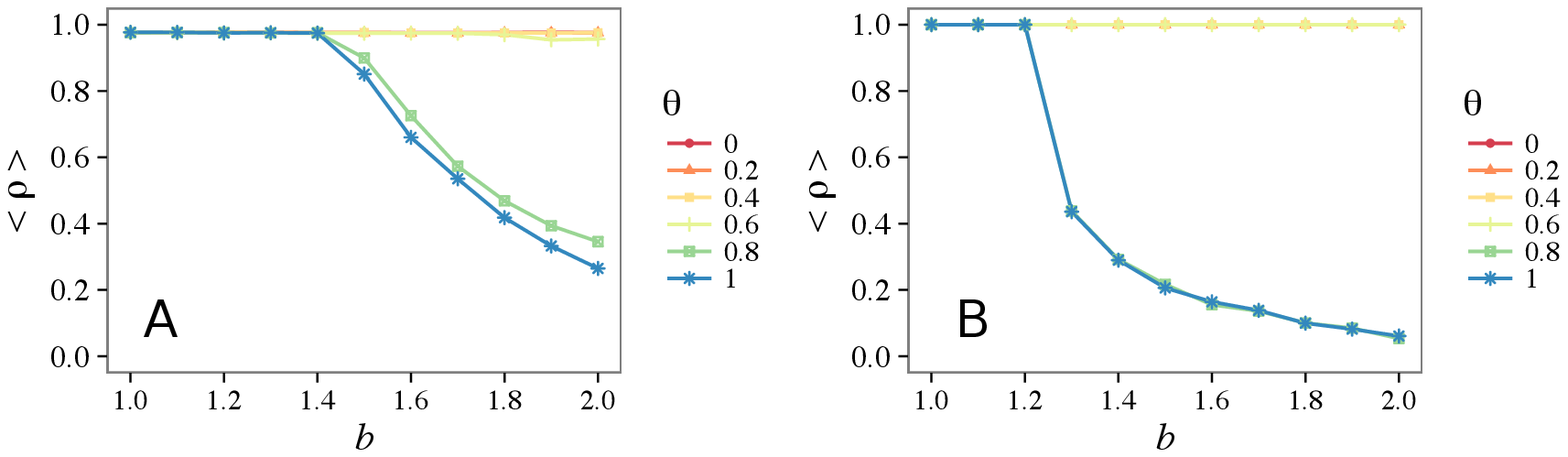}
\caption{ER monoplex network. {\bf A}: $z=4$; {\bf B}: $z=16$. Average final fractions of cooperators $\rho$ as \mbox{a function of} $b$ for different values of $\theta$ --- {REP evolution rule.}}
\label{monoER_DNV}
\end{figure}

\begin{figure}[H]
\centering
\includegraphics[width=13cm,angle=0]{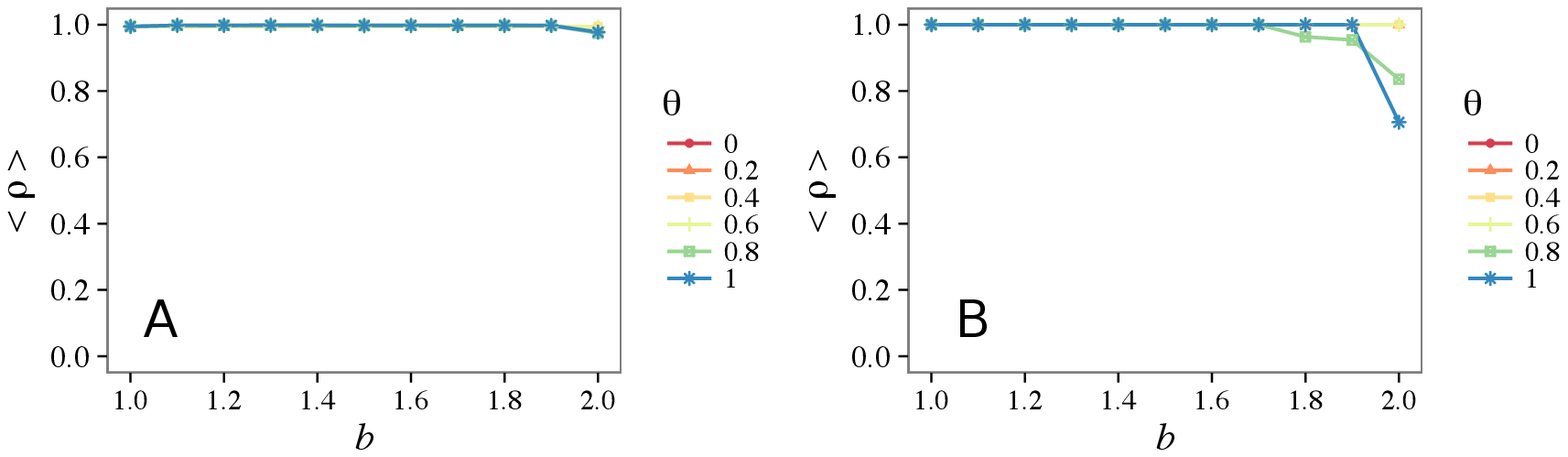}
\caption{BA monoplex network. {\bf A}: $z=4$; {\bf B}: $z=16$. Average final fractions of cooperators $\rho$ as \mbox{a function of }$b$ for different values of $\theta$ --- {REP evolution rule. Notice that all the graphics for the \mbox{six different} values of $\theta$ coincide almost everywhere.}}
\label{monoBA_DNV}
\end{figure}

As Figures~\ref{monoER_DNV} and ~\ref{monoBA_DNV} show, the fact that a defector also contributes to vigilance pushes cooperation dramatically both in ER and BA monoplex networks, having full cooperation in BA monoplex networks for all values of $\theta$ and $b$.

Now, we show the results for a square lattice, where the effect is expected to be magnified with respect to the remaining topologies. Actually, as shown in Figure~\ref{uila}{\bf A}, in this topology, cooperation is mostly hindered.

\begin{figure}[H]
\centering
\includegraphics[width=15cm,angle=0]{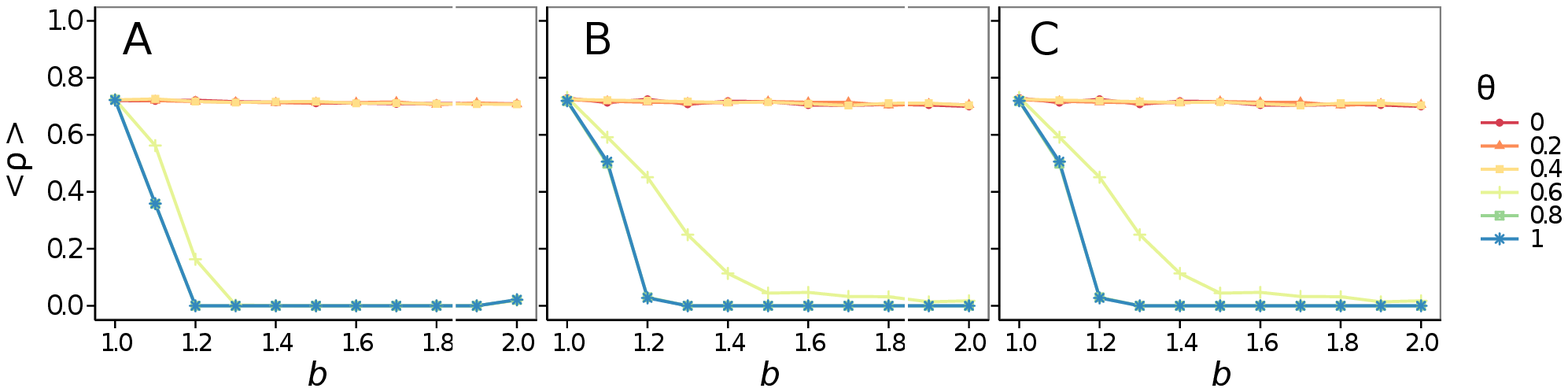} \\
\caption{Lattice, vigilance independent from strategy. \textbf{A}: initial probability to be vigilant 0.001; \textbf{B}: initial probability to be vigilant 0.05; \textbf{C}: initial probability to be vigilant 0.45. Average final fractions of cooperators $\rho$ as a function of $b$ for different values of $\theta$ --- {REP evolution rule. In these graphs, the curves for $\theta=0.8$ and $1$ assume the same values.}}
\label{def_vig}
\end{figure}

In Figure~\ref{def_vig} we see that already a very small probability $\varrho_0$ of being an initial vigilant ({\bf A}) helps cooperation to invade the population, already for not-too-high vigilance ($\theta\lesssim0.5$), \mbox{and almost the same for} $\varrho_0\gtrsim0.05$ ({\bf B} and {\bf C}). In addition, for $\theta=0.6$, the final cooperator density does not vanish even at higher values of $b$. This is, of course, an expected result, since allowing more individuals to activate as vigilant ones decreases much more the average temptation of every player, according to the Equation~(\ref{eq:vig}). This outcome holds also in different topologies.

\ 

\section{Discussion and conclusions}
\label{discuss}

The model of vigilance firstly presented in Ref.~\cite{per16}, and further developed here, treats the pro-social effect due to the control (be it real or just perceived) by peers as a decreasing of the temptation to defect: the more neighbors watch out for the behavior of a subject, the less is the probability that the latter adopts a selfish strategy. Even though the preliminary results of this approach turned out to be promising, before considering it a viable way to describe this phenomenon, it was necessary to test its full validity. Therefore, in this paper, we have aimed to ascertain that the main features of the model are basically robust: that is, we verified that, through the mechanism of vigilance proposed here, cooperation is actually fostered for broad values of the parameters at stake and in different environmental configurations. In particular, we showed that the beneficial influence of the vigilance works in more realistic configurations, allowing us to hypothesize that what has been repeatedly observed in experiments and field observations can be actually explained as a smaller temptation to defect in presence of controllers.

The results, which, in our opinion, allow us to consider the model realistic {and particularly useful} are the following:

\begin{itemize}
\item Vigilance needs the small-world effect (the presence of short-cuts connecting individuals physically far away from each other) to be efficient in fostering cooperation: {indeed, in} regular lattices, Figure~\ref{uila}, it does not help, and the small-world property is ubiquitous in most real social systems {(only the smallest communities can be modeled by complete graphs, and Euclidean topologies are even more uncommon in human societies)}.
\item Vigilance works not only when the individuals update their strategy by means of an essentially evolutionary rule (REP), but also when they evolve through more typically ``social'' mechanisms as pure imitation (at least on ER networks); moreover, considering the mixed rule, \mbox{which takes} into account the intrinsic non-strategic component of humans' decision making processes, \mbox{we found that} the cooperation can tolerate the influence of irrationality only when this is low ($q<0.5$), coherently with the results of Ref.~\cite{vil14}.
\item{Concerning again the update rule, it is worth stressing that, in heterogeneous networks (scale-free), vigilance is beneficial for cooperation only with replicator update, whilst with strategic imitation (UI) the presence of hubs appears to be detrimental for the emergence of pro-social behaviors}.
\item The results do not depend sensitively on the initial conditions (at least in {heterogeneous topologies}): this is a fundamental feature of the model since it is usually hard to determine the initial conditions for real social systems; on the other hand, in complete graphs \mbox{(i.e., in mean-field }approximation), this is not true, but only small human communities can be described in this way, and, in such cases, different dynamical mechanisms are at work~\cite{gua12}.

\end{itemize}

Therefore, we can state again the main result of this work: to confirm the reliability of the model and its potentiality.
Of course, further investigations are needed to validate definitively the model --- in particular, experiments explicitly aimed to check if this peculiar kind of phenomenon (decreased temptation in a PDG) actually takes place when subjects play in the laboratory. These kinds of studies are already planned for future work.

\ 

\section*{Acknowledgments}

M.P. acknowledges support from the project H2020 FET OPEN RIA IBSEN/662725 and the Institute of Physics of Cantabria (IFCA-CSIS) for providing access to the Altamira supercomputer. D.V. acknowledges support from H2020 FETPROACT-GSS CIMPLEX Grant No. 641191, which funded the publication of this article and the CNR (Consiglio Nazionale delle Ricerche) for the Short Term Mobility Program 2016, which funded his stay at the Universidad Carlos III de Madrid where this work was accomplished.

\

\subsubsection*{Author contributions}

M.P. created the original model and accomplished the simulations. D.V. conceived the model generalizations and conducted the theoretical analysis. Both authors wrote and revised the manuscript. The authors declare no conflict of interest.

\end{document}